eJournal interface can influence usage statistics: implications for libraries,

publishers, and Project COUNTER.

[Final manuscript accepted for publication in JASIST. Updates current to July 27, 2005]

Philip M. Davis
Life Sciences Librarian
Albert R. Mann Library
Cornell University, Ithaca NY 14853-4301
pmd8@cornell.edu
607-255-7192 (phone)

Jason S.Price
Life Science Librarian
Libraries of the Claremont Colleges
640 N College Ave. Claremont, CA 91711
jprice@libraries.claremont.edu
909-621-8437

# Abstract

The design of a publisher's electronic interface can have a measurable effect on electronic journal usage statistics. A study of journal usage from six COUNTER-compliant publishers at thirty-two research institutions in the United States, the United Kingdom and Sweden indicates that the ratio of PDF to HTML views is not consistent across publisher interfaces, even after controlling for differences in publisher content. The number of fulltext downloads may be artificially inflated when publishers require users to view HTML versions before accessing PDF versions or when linking mechanisms, such as CrossRef, direct users to the full text, rather than the abstract, of each article. These results suggest that usage reports from COUNTER-compliant publishers are not directly comparable in their current form. One solution may be to modify publisher numbers with 'adjustment factors' deemed to be representative of the benefit or disadvantage due to its interface. Standardization of some interface and linking protocols may obviate these differences and allow for more accurate cross-publisher comparisons.

# Introduction

The counting and reporting of electronic usage statistics has affected the ways in which both publishers and librarians assess the value of information (Mercer, 2000; Rous, 2004). For the publisher, usage is an indication of how the scholarly community values its journals, and may thereby influence pricing. Librarians in turn may need to justify the funds they allocate to purchasing certain journals in terms of their usefulness to the community. A low annual cost per article download may provide a validation for continuing a subscription – the opposite may provide some rationalization for cancellation.

Justifying price based on utility appears to be on the forefront of many publishers' minds. In the summer of 2004, the United Kingdom House of Commons Parliamentary proceedings on Scholarly Journals confirmed this position for several top British publishers. In response to a question asking how publishers were able to explain historically high annual journal price increases, Chrispin Davis, the CEO of Reed Elsevier responded,

> "The biggest single factor is usage. That is what librarians look at more than anything else and it is what they [use to] determine whether they renew, do not renew and so on. We have usage going up by an average of 75 per cent each year. In other words, the cost per article download is coming down by around 70 per cent each year. That is fantastic value for money in terms of the institution, so I would say that [usage] is the single biggest factor." (House of Commons, 2004)

This response illustrates a dramatic change in the rationale for journal pricing. It is no longer a matter of absolute journal costs, but of relative ones, and as long as a publisher can demonstrate high numbers of article downloads, it can justify high prices. Even the most expensive journal in the sciences, *Brain Research,* can be made to look like a good value for libraries if it can demonstrate a sufficiently high number of article downloads each year.

## Literature Review

Project COUNTER was formed in 2002 by publishers, vendors and librarians to develop an internationally recognized standard for the way electronic resource use is counted and reported to customers (Project COUNTER, 2005a). One of the many benefits for librarians is the ability to compare the performance of similar journals or journal packages across publishers.

The function of HTML as a medium for browsing and PDF as a medium for printing was first rigorously documented in the SuperJournal project in the late 1990s, (Eason et al., 2000) and has been confirmed anecdotally ever since. In the early 2000's, a study tracing the use of fourteen electronic journals hosted with HighWire Press quantified the likelihood of readers downloading first the HTML and then the PDF version of the same article (Institute, 2002). However, since the current COUNTER standard allows publishers to simply report the total number of full text article requests (HTML and PDF downloads added together), a publisher that provides only one format may not be comparable to another publisher providing both formats, despite of the fact that both publishers are COUNTER-compliant.

In 2003, Davis and Solla (2003) reported a strong linear relationship between the number of article downloads and the size of a user population for twenty-nine journals published by the American Chemical Society. These findings were confirmed again by Davis in 2004 for sixteen research institutions in the United States, United Kingdom and Sweden with journals published by HighWire Press (Davis, 2004). His results indicated that the relationship between fulltext downloads and the size of a user population is internally consistent within a publisher's set of journals irrespective of the subject, size, or popularity of its journals. This relationship is also consistent over time and across institutions, meaning that the results of a large institution like Cornell University can be regarded as a representative sample of the universe of research institutions. Yet, in comparing the results for HighWire journals to his previous results for ACS journals, Davis discovered that this relationship between fulltext downloads and size of a user population was not consistent across these publishers. As a possible cause, he postulated that some "interface effect" may be responsible for the difference. He then described the logical shortcomings of blindly attributing differences across publishers to differences in their interfaces. He wrote:

"In a strict sense, it is impossible to directly estimate an interface effect when comparing different content on different sites. Such an experiment would require the same journals to be hosted by different providers so that the effect of an interface could not be confused with content. Consequently, estimating the interface effect can only be done by controlling for all other possible contributing factors." (Davis, 2004)

## Study rationale

This study attempts to measure the effect of the publisher's interface on usage statistics for six COUNTER-compliant publishers. If each publisher is counting the same way, then one should see, in theory, similar journal usage patterns across all six publishers. The usage pattern sought in this study is not the total number of downloads, but the ratio of PDF to HTML downloads. To control for different content across publishers, the authors then compare the usage pattern for a journal hosted simultaneously on two different platforms. In essence, this study poses two major questions:

*Q1. Does the ratio of PDF-to-HTML downloads differ between publishers?*

*Q2. Does this difference hold when controlling for content?*

In order for collection developers to make impartial value comparisons among journal content provided by different publishers, it is necessary to remove strong biases due to publisher interface design. This paper provides evidence that such biases exist in some publisher interfaces and suggests a method which can be used to mitigate their effects.

# Methods

*Q1. Does the ratio of PDF-to-HTML downloads differ between publishers?*

Journal usage statistics for six Counter-compliant publishers were gathered for Cornell University for the year 2004. Only publishers that provided Project COUNTER usage report 1a (separating fulltext article requests into PDF and HTML requests) were used. These publishers included: The American Chemical Society (ACS), Blackwell Publishing, HighWire Press, Nature Publishing Group, Oxford University Press (OUP) and Wiley Publishers. Journals that function as news sources (e.g. ScienceNow and News@Nature) were removed from the dataset. Some publishers host journals (or journal archives) that are only released in one format (HTML or PDF). As a consequence, all titles reporting either zero HTML or PDF downloads were removed from the dataset. These exclusions ensured that only scholarly journals with meaningful PDF to HTML ratios were included in the dataset. Out of a possible 1590 titles, 818 remained for analysis.

An univariate General Linear Model was used with the $\log_{10}$ PDF-to-HTML ratio as the dependent variable, and the publisher as the fixed factor. Raw PDF to HTML ratios were log transformed to satisfy the assumption that the data are normally distributed – a prerequisite for GLM analysis. In addition, the raw PDF and HTML download values were used as covariates to account for differences in popularity among journals. Estimated marginal mean differences were compared for each publisher pair, using the Bonferroni adjustment for multiple comparisons.

*2. Does this difference hold when controlling for content?*

To determine whether identical content can demonstrate different usage patterns on two publisher platforms, total PDF and HTML downloads were gathered for *The Embo Journal* from thirty-two participating research institutions in the United States, the United Kingdom and Sweden (see Acknowledgements for a list of participating institutions). *The Embo Journal* is hosted simultaneously on both the Nature and HighWire platform. A paired t-test was used to compare PDF-to-HTML ratio across these two platforms.

The statistical tests used in this paper pertain to the six publishers involved in the study and do not necessarily reflect all scientific publishers. The detection of significant differences among these publishers, however, would refute the notion that all COUNTER-compliant publishers are reporting comparable usage numbers.

## Results

The mean journal PDF-to-HTML ratio differs across publishers, ranging from a low of about 1:1 for HighWire Press to as high as almost 20:1 for Wiley (Figure 1.)

A scatter plot of HTML vs. PDF downloads for each journal shows a strong relationship between these two variables across all journals ($R^2$ = .742, Figure 2). The six publisher regression lines for these data are presented without data points (as Figure 3), illustrating that there are differences in the slopes (ratio) and y-intercepts among these publishers.

The results from the General Linear Model analysis revealed a significant difference in the PDF-to-HTML ratio across publishers (P<0.001), even after controlling for the number of downloads for each journal (Table 1).

Table 2 reports pairwise comparisons of publisher mean log PDF-to-HTML ratios. For example, the ACS had a significantly higher mean ratio than all other publishers except Wiley, and Wiley had a significantly higher mean ratio than the remaining four publishers. Of the fifteen possible pairs of publishers, ten of them differ significantly at the .05 level.

## Testing Identical Content on Different Publisher Platforms

Our comparison of identical content hosted on two different platforms demonstrated significantly different usage patterns. The box-plot (Figure 4) illustrates the difference in the PDF-to-HTML ratios for *The Embo Journal* hosted on the Nature and HighWire platforms. For the thirty-two participating institutions, the mean PDF-to-HTML ratio for this journal was 1.23 for the Nature interface and 0.59 for the HighWire interface. A paired t-test comparing this ratio across 22 of the 32 institutions that provided data for both interfaces reveals a clearly significant difference (P<.001). Thus, even when controlling for content, a journal hosted on the Nature platform can result in a PDF-to-HTML ratio that is twice that of the HighWire platform.

# Discussion

The internal consistency of PDF-to-HTML downloads for journals hosted on a publisher's platform, as revealed by strong linear relationships within publishers (Figure 2), suggests that user behavior is not affected by factors such as subject scope, frequency, or popularity of a journal. This result is consistent with earlier studies (Davis & Solla, 2003; Davis, 2004).

The PDF-to-HTML ratio varied considerably across publishers, even after controlling for popularity (total PDF and HTML downloads), suggesting that a publisher's interface can have a measurable effect on its usage statistics. Furthermore, interface can be an explanatory cause in and of itself, as demonstrated by significantly different PDF-to-HTML use ratios of identical content on the HighWire and Nature platforms (Figure 4).

While the HighWire and Nature interfaces to *The Embo Journal* are visually similar, a comparison between the two illustrates a striking difference in how a user navigates to the PDF version of an article. A researcher starting at the table of contents page on the HighWire interface is required to first download the article in HTML. For the Nature interface, a direct link to the PDF is available directly from the table of contents (Figure 5). This pathway dissimilarity is a likely explanation for the significantly lower ratio of PDF-to-HTML downloads for *The Embo Journal* hosted on the HighWire platform. As a result of this study, HighWire Press announced that a link to the PDF version will be included in all new issues of its journals beginning July 2005. Oxford University Press (which also publishes on the HighWire platform), subsequently announced that it will begin retrospective conversion of its interface to include PDF links from past issues.

The concept of a publisher's "interface" can be much broader than the platform alone. There are many different ways to get to a publisher's content, and the journal's table of contents is but one method. Direct links from indexes such as PubMed, SciFinder Scholar, or Google Scholar are frequent sources of referral to a publisher's content, as are direct article-to-article links facilitated by protocols such as CrossRef. In addition, a publisher may register a preference with CrossRef to direct traffic to its Abstract, HTML or PDF version. HighWire directs all external CrossRef links to the HTML version, while Blackwell, Nature and Wiley direct external links to the abstract page, and ACS directs external links to a citation page. Such publisher preferences may have an effect on the on the number of fulltext downloads and is likely to be a partial explanation of the differences in the PDF-to-HTML ratio reported in this study.

Lastly, journal archives may have a small effect on the PDF-to-HTML ratio, since many journal archives only provide PDF versions of articles. Three of the publishers in this study (ACS, Wiley, Nature) separate out their archives when reporting journal usage (which were subsequently removed from the dataset prior to analysis), while Blackwell, HighWire and Oxford do not. These inconsistencies in publisher reporting structure are likely to have a minimal effect on the results given that most journal usage represents downloads of recently published articles. According to data from HighWire press, the majority of article downloads for *The Embo Journal* take place within three-months of an article being published (Richardson, 2001).

## Implications

One of the key rationales for the creation of the Project COUNTER standard was to enable unbiased usage comparisons across publishers. The results of this research suggest that this may not be possible. While the findings of this study cannot be statistically generalized to all forty COUNTER-compliant publishers, the fact that there is a lack of consistency among the six that could be assessed suggests a general problem of standardization. Project COUNTER was not created to dictate how a publisher hosts and makes content accessible, and any attempt to standardize publishers' interfaces would be unanimously resisted. On the other hand, librarians need usage reports that can be deemed credible and comparable. How can this be done when usage statistics are heavily influenced by a publisher's interface?

One solution may be to modify publisher numbers with 'adjustment factors' that represent the benefit or disadvantage from its interface. Since Cornell's numbers have been demonstrated to be consistent across time and across institutions, each institution could rely upon a generally-accepted set of adjustment factors, as presented in Figure 1. Further study needs to be devoted to how these ratios can be used to provide optimal adjustment factors that reflect the value of the content as accurately as possible. Commencing January 2006, all COUNTER-compliant publishers will be required to make separate HTML and PDF counts available as part of the required Journal Report 1 (Project COUNTER, 2005b). This will allow follow-up studies to include all (rather than a subset) of COUNTER-compliant publishers.

While this study was able to measure the effect of individual publisher interfaces on usage data, it was unable to discern exactly *which features* were responsible for those differences. The scientific process of measuring the interface effect in this study was done by eliminating all other possible explanatory factors. A more direct approach would provide greater insight as to which aspects of an interface are the most important. For example, one could compare the use ratios *before* and *after* a publisher changed its interface, such as when Oxford University Press adds a link to the PDF version from its past table of contents pages. This could help publishers develop superior interfaces that enhance rather than impede a researcher's access to the literature. On the other hand, it is entirely possible for a publisher to optimize its interface to maximize the total number of article downloads. In an environment that requires justifying price per download, this may be an understandable goal.

## Acknowledgements

The authors wish to thank the following institutions of higher education for participating in this study by sharing their usage data: Brown, CalTech, Columbia, Cornell, Duke, Harvard, Iowa State, Johns Hopkins, Kings College (UK), Lund (Sweden), Northwestern, NYU, Penn State, Rutgers, U. Alabama, U. Bristol (UK), U. Chicago, U. Connecticut, U. Georgia, U. Glasgow (UK), U. Leicester (UK), U. Liverpool (UK), U. Massachusetts, U.

Michigan, U. Nevada, U. Notre Dame, U. Rochester, U. Tennessee, U. Virginia, U. Washington, U. Wisconsin, Yale.

The authors also wish to thank John Sack at HighWire Press and Jack Ochs at the American Chemical Society for help in understanding the broad functions of a publishers interface, and to Bill Walters and Suzanne Cohen for their scrupulous review and suggestions.

# Table and Figures

**Figure 1.  Mean ratio of PDF to HTML article downloads varies across publishers.  Cornell University downloads, 2004**

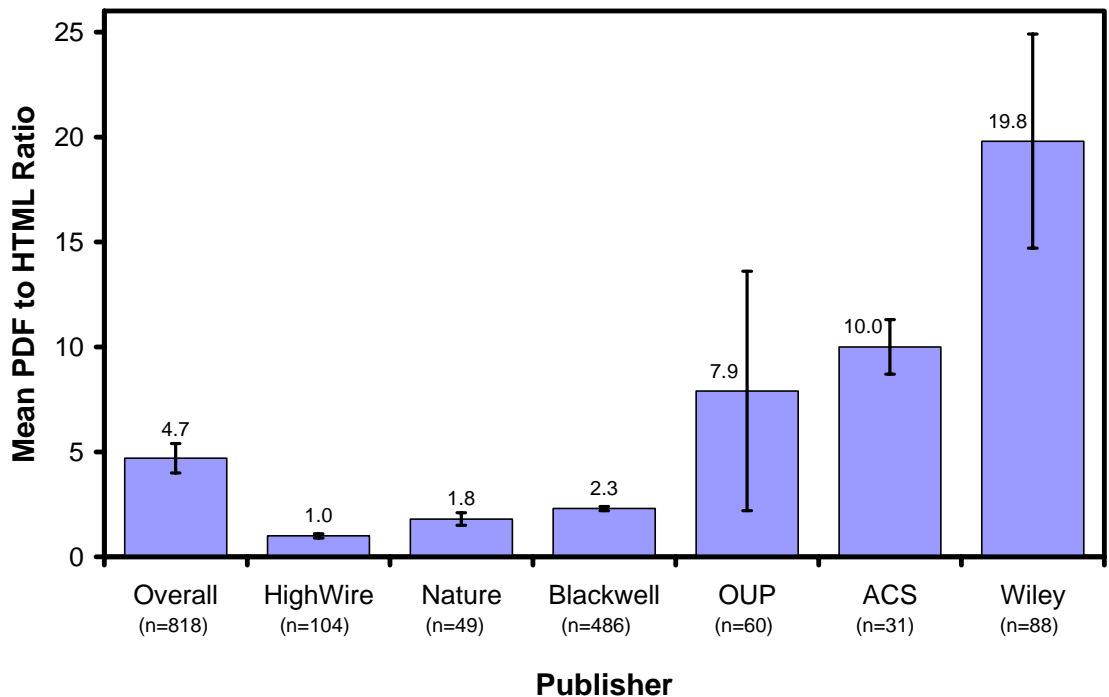

**Figure 2. Scatter plot of journal use for all six COUNTER-compliant publishers.**

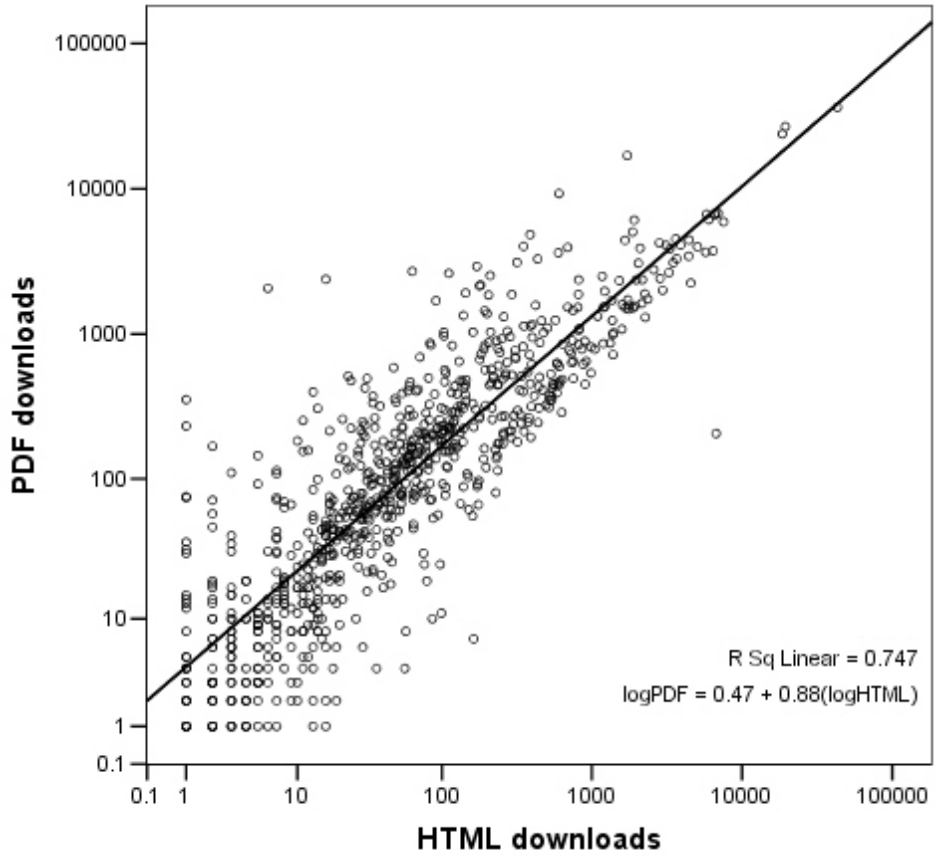

Notes: Each point represents the total HTML and PDF downloads for a subscribed journal at Cornell University for 2004.

**Figure 3. Regression lines for each COUNTER-compliant publisher (data points removed for clarity).**

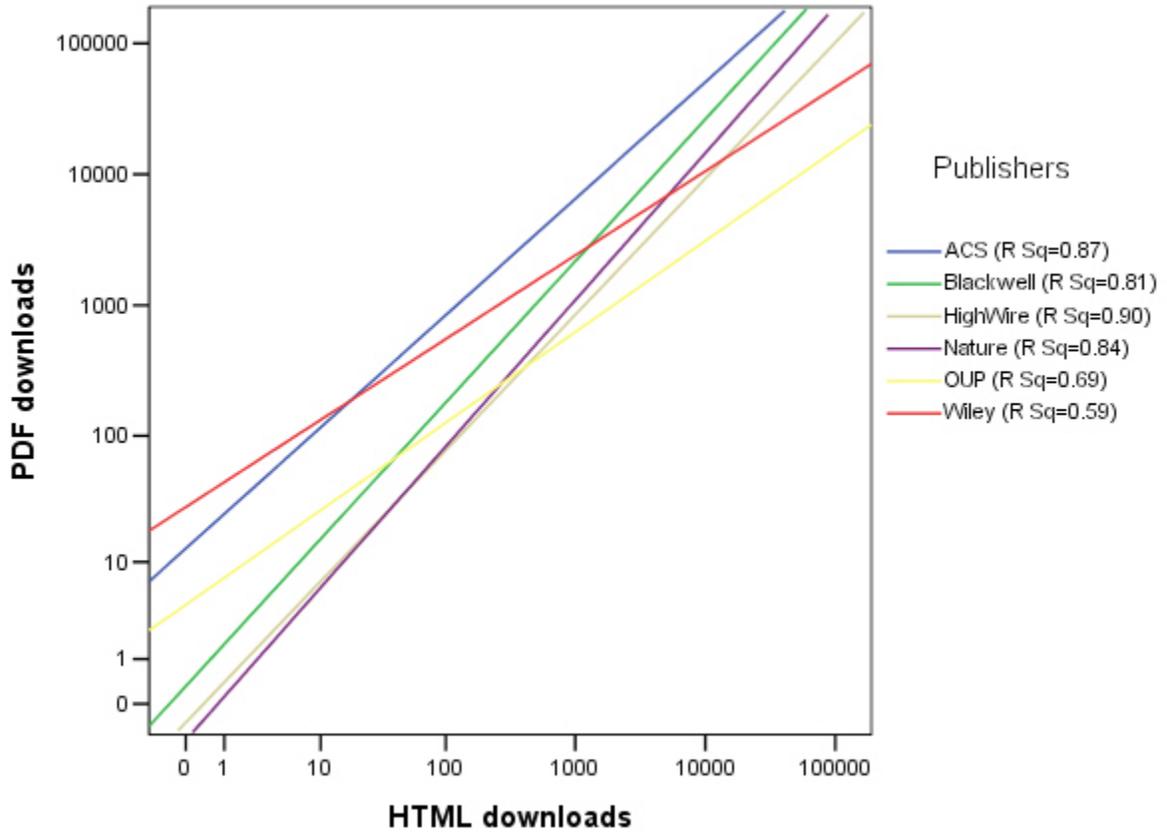

Notes: All journals exhibiting either zero HTML or PDF downloads were excluded from the analysis to ensure that only titles that are published in both formats are compared.

# Table 1. Testing the effect of Publisher on log PDF-to-HTML ratio, holding downloads constant.

Dependent Variable: log_ratio

| Source | df | F | P-value |
|---|---|---|---|
| Corrected Model | 7 | 83.596 | <0.001 |
| Intercept | 1 | 251.830 | <0.001 |
| HTML | 1 | 26.224 | <0.001 |
| PDF | 1 | 29.480 | <0.001 |
| Publisher | 5 | 86.665 | <0.001 |
| Error | 810 | | |
| Total | 818 | | |
| Corrected Total | 817 | | |

Corrected Model R Squared = .419 (Adjusted R Squared = .414)

**Table 2. Pairwise comparisons of publisher means (log PDF to HTML ratio) across all journals.**

Dependent Variable: log PDF:HTML ratio

| Publisher (I) | Publisher (J) | Mean Difference (I-J) | P value[a] |
|---|---|---|---|
| ACS | Blackwell | **.543*** | **< .001** |
| | HighWire | **.839*** | **< .001** |
| | Nature | **.762*** | **< .001** |
| | OUP | **.669*** | **< .001** |
| | Wiley | -.151 | = .846 |
| Blackwell | HighWire | **.296*** | **< .001** |
| | Nature | **.220*** | **< .001** |
| | OUP | .126 | = .171 |
| | Wiley | **-.693*** | **< .001** |
| Highwire | Nature | -.076 | = 1.000 |
| | OUP | -.170 | = .064 |
| | Wiley | **-.990*** | **< .001** |
| Nature | OUP | -.094 | = 1.000 |
| | Wiley | **-.913*** | **< .001** |
| OUP | Wiley | **-.819*** | **< .001** |

Based on estimated marginal means
  *. The mean difference is significant at the .05 level.
  a. Adjustment for multiple comparisons: Bonferroni.

Notes: Comparisons were made between all 15 possible pairs of publishers, reported in alphabetical order.  Negative values indicate that the mean ratio of the publisher in the second column was greater than that in the first.

**Figure 4. Ratio of PDF-to-HTML for *The Embo Journal* hosted on two different publisher platforms for thirty-two research institutions.**

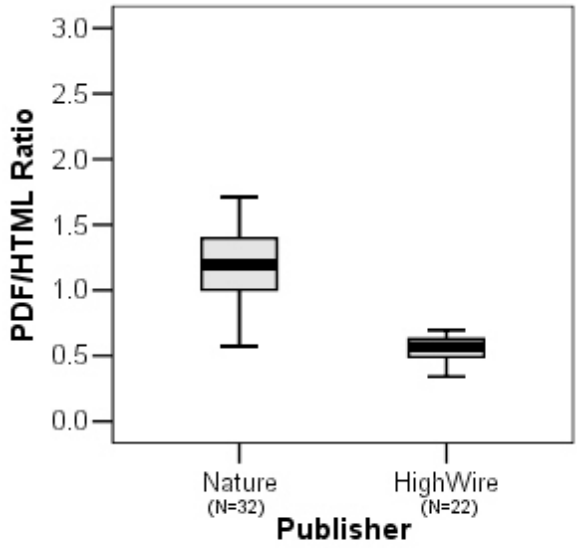

Notes:
Significant at P<.001 (2-tailed paired t-test).

**Figure 5. HighWire versus Nature TOC interface for the same article published in *The Embo Journal*.**

**4a. HighWire**

☐ María A Blasco

**Mice with bad ends: mouse models for the study of telomeres and telomerase in cancer and aging**

EMBO J. 2005 24: 1095-1103; advance online publication, March 10, 2005; doi:10.1038/sj.emboj.7600598.

[Abstract] [Full Text]

**4b. Nature**

**Mice with bad ends: mouse models for the study of telomeres and telomerase in cancer and aging**

María A Blasco

*The EMBO Journal* (2005) **24,** 1095–1103, doi: 10.1038/sj.emboj.7600598

| Abstract | Full text | PDF (499K) |

Published online: 10 March 2005

**Subject Categories:** Chromatin & Transcription | Molecular Biology of Disease

-------

Notes: In order to access PDF from HighWire interface, a user is required to download the article first in HTML (Full Text).